\documentclass[10pt,a4paper]{article}
\usepackage{amsmath}
\usepackage{amsthm}
\usepackage{epsfig}
\usepackage{latexsym}
\usepackage{xcolor}
\usepackage{pdflscape}
\numberwithin{equation}{section}
\usepackage{graphicx}
\usepackage{dcolumn}
\title{\Large \bf Dynamical phase diagrams of a love capacity  constrained prey-predator model   }
\author{  P. Toranj Simin$^1$,  G.R. Jafari$^{1,2}$ \\M. Ausloos$^{3,4}$, \\C. F. Caiafa$^{5,6}$, \\F. Caram$^7$,
\\A. Sonubi$^8$, A. Arcagni$^8$, S. Stefani$^8$,
}

 \date{
$^1$ Department of Physics, Shahid Beheshti University, G.C., \\ Evin, Tehran 19839, Iran \\
$^2$ Center for Network Science, Central European University, H-1051, Budapest, Hungary \\
$^3$ School of Business, University of Leicester, University Road, Leicester  LE1 7RH, United Kingdom \\
$^4$ GRAPES\footnote{Group of Researchers for Applications of Physics in Economy and Sociology}$\;$,
rue de la Belle Jardini\`ere 483, \\B-4031, Angleur, Belgium\\
$^5$ Psychological and Brain Science, Indiana University, Bloomington, USA\\
$^6$ Instituto Argentino de Radioastronomía, CCT La Plata, CONICET, Argentina. \\ 
$^7$ Laboratorio de Redes y Sistemas M\' oviles, FI-UBA, Argentina\\
$^8$ Scuola di Economia e Statistica, Department of Statistics and Quantitative Methods,\\
Universit\`a degli Studi di Milano-Bicocca, I-20126 Milano, Italy\\
}
\begin{document}
 \maketitle


\begin{abstract} 
{\it 
One interesting question in love relationships is: finally, what  and when is the end of this love relationship?
Using a prey-predator Verhulst-Lotka-Volterra (VLV) model we imply cooperation and competition tendency between people in  order  to describe a "love dilemma game". We select the most simple but immediately most complex case for studying the set of nonlinear differential equations, i.e. that  implying three persons, being at the same time prey and predator. We describe four different scenarios in such a love game containing either a one-way love or a love triangle. Our results show that it is hard to love more than one person simultaneously. Moreover,   to love several people simultaneously is an unstable state. We find some condition in which persons tend to have a friendly relationship and love someone in spite of their antagonistic interaction. We demonstrate the dynamics by displaying flow diagrams.
}

\end{abstract}
\section{INTRODUCTION}\label{introduction}
 
Empirical studies and theoretical modeling of interacting agents have been the subject of a large body of recent research in 
statistical physics and applied mathematics \cite{DoroMendes}.  No need to recall (an exhaustive list of references would be impossible neither to print nor read) the  flurry of papers on the "prisoner's dilemma game" since it was proposed by  Flood and Dresher in 1950  \cite{RapoportChammah,Poundstone}.  Nowadays, agents are considered to behave on networks rather than on lattices.  Their "opinion evolution" can be imagined to occur according to complex algorithms, but also basing their evolutive behavior on non linear differential equations. Let it be recalled that the Lotka-Volterra  differential equation set  for preys-predators  \cite{volt}  has been used in various ways to model many types of complex systems, with competitive or cooperation aspects. For instance,  outside demography or biology, effects of competition on growth,  pertinent in economy and opinion formation 
  \cite{PhA389.10NKVZIDMAideol,ACSmigration}   taking into account a "market capacity" \cite{v1,v2} has been studied in  \cite{caram1,caram2,sonubi1}.

As recently pointed out \cite{NDPLS}, dynamic system theory is useful for describing complex behavior of humans. The usefulness of non linear differential equations is hereby used for outlining possible "love story" scenarios. The classical Lotka-Volterra equations \cite{volt} for  prey-predator population evolution is taken as a model for one of the most common human behavior: love. The psychological conditions of a triad of potential lovers is described through an adjacency matrix. The mathematical analysis follows ideas presented elsewhere for economic-financial markets \cite{caram1,caram2,sonubi1}, introducing the notion of maximum love capacity, as in Verhulst logistic equation\cite{v1,v2},  that is, limited resources for a population in a country.

The purpose of the described work here below touches upon a  classical aspect of human nature, described by many but rarely touched upon in terms of complex system theories, - yet quite useful in quantitative social sciences, including psychology and cognition \cite{NDPLS}. Nevertheless, it is fair to point to a pioneering work on the matter by Strogatz \cite{Strogatz}.  He considers the evolution of love in a couple. We add that such an evolution depends in an endogenous way of the possible other partners. Moreover,  it is reasonable to  consider an exogenous constraint:  a "love capacity" concept can also be introduced.

In Sect. \ref{method}, we present and develop a generalized Verhulst-Lotka-Volterra model in order to describe the "love states" of three interacting persons. In Sect. \ref{results}, we analyze the  four possible interaction schemes, searching for  steady states through a fixed point and stability analysis. We present  the non trivial dynamical phase diagrams, depending on initial love conditions of the partners.  In Sect. \ref{conclusions}, we point to the qualitative "love story" aspects mimicked by our  quantifying non-linear dynamics model, - and suggest extensions.

\section{METHOD}\label{method}

The classical Lotka-Volterra equations or the predator-prey model equations are used to describe the dynamics of  a  population  made of two species which interact, one as a  predator and the other as a prey. The size $s_i$ of populations  ($i=$ 1 or 2) changes according to the set of equations 
\begin{equation} \dot{s_{i}} = \lambda_{i} s_{i}- \zeta_{i} s_{i} s_{j} 
\end{equation} 
where $\lambda$'s and $\zeta$'s are system parameters.

On the other hand, the simple logistic growth equation also known as the Verhulst equation describes the time evolution of a single species. Interestingly , and realistically, this equation considers the lack of resources in the growth of the population of a species according to the equation 

\begin{equation} \dot{s_{i}} =  s_{i} (a - b s_{i} )
\end{equation}	
where the parameter $b$ is related to the available resources  for a species. 

One can combine these two aspects of  a  population growth of species to find a generalized model which describes the evolution of the population of several species, on a "market" with finite resources. The combination of these equations leads to the generalized Verhulst-Lotka-Volterra model  \cite{caram1,caram2,sonubi1}

\begin{equation} \dot{s_{i}} = \alpha_{i} s_{i}(\beta_{i} - s_{i})
 - \sum_{i \neq j} \gamma(s_{i},s_{j}) s_{i} s_{j}, 
 \quad \quad \quad i=1,\ldots,n,
\end{equation}\label{eqbasic}

In the present case,  $s_i$ is the "size of the love" of the agent \emph{i}, such that $0<s_i<1$;  $\alpha_i$ is the growth rate for the love when there is no interaction; $\beta_i$ is the maximum capacity of the love for the agent \emph{i}.    
According to psychological aspects, the maximum amount of "love capacity" $\beta$  is supposed to be  such that
\begin{equation}
 \nonumber \beta_{i} = \beta - \sum_{j\neq i} s_{j}, 
\end{equation}
similar to market constraints in economy and finance \cite{caram1,caram2,sonubi1}.
For simplicity, we may assume  that all the lovers have the same growth rate $\alpha_i=1$ and also  assume  that the love capacity is  $\beta=1$.
 
The interaction function in Eq. (2.3)  is defined by 
\begin{equation}
\gamma(s_i,s_j)\equiv \Gamma_{i,j} \; exp{[-(\frac{s_i-s_j}{\sigma})^2]}.
\end{equation}
This Gaussian term expresses the effect of the difference between the strength of the love between two agents. This term notices some life reality. the interaction function $\gamma(s_i,s_j)$ is the largest if the strength of the lover $i$ and that of  $j$ are similar, while   $\gamma(s_i,s_j) \approx 0$ if the strengths are quite different, - though not neglected. 

 The parameter $\sigma$ is positive and scales the  intensity distribution of the "lover strengths". For simplicity, we assume $\sigma=1$.

The parameter $\Gamma_{i j}$ can be considered as an element of a matrix $\Gamma$ which specifies various scenarios on the types of lover's interactions.

Here we define four different $\Gamma$-matrices each of which characterizes different love scenarios among three lovers $A_1$, $A_2$ and $A_3$.
The possible $\Gamma$-matrices for describing one-way love are
\begin{equation}
\begin{split}
\Gamma_1=\left( \begin{array}{ccc}
0 & -1 & -1 \\ 
1 & 0 & 0 \\ 
1 & 0 & 0 \end{array}
\right), \; 
 \Gamma_2=\left( \begin{array}{ccc}
0 & 1 & 1 \\ 
-1 & 0 & 0 \\ 
-1 & 0 & 0 \end{array}
\right)
\end{split}
\end{equation}
$\Gamma_1$ represent a situation that $A_1$ is not interested in $A_2$ and $A_3$, but $A_2$ and $A_3$ care about $A_1$ and wish $A_1$ as a lover. This can be imagined as a situation in which two individuals fall in love with a (famous) person, - who {\it a priori} does not care.

In contrast, $\Gamma_2$ expresses that $A_1$ falls in love with both $A_2$ and $A_3$, but $A_2$ and $A_3$ do not care about $A_1$.

We also describe a famous situation in relationships called "love triangle". Two main forms of love triangle have  to be  distinguished:  there is the rivalrous triangle, where the lover is competing with a rival for the love of the desired, and the split-object triangle, where a lover has split  some  attention between two love objects".

Thus,  we consider $A_1$ is interested in both $A_2$ and $A_3$, and also $A_2$ and $A_3$ care about $A_1$ and wish love;  $A_2$ and $A_3$ do not  "know" each other, such that there is no  interaction between $A_2$ and $A_3$. The possible $\Gamma$-matrix for expressing such a love triangle game is
\begin{equation}
\Gamma_3=\left( \begin{array}{ccc}
0 & -1 & -1 \\ 
-1 & 0 & 0 \\ 
-1 & 0 & 0 \end{array}
\right)
\end{equation}
In contrast, one may consider a situation in which  $A_1$ does not like  $A_2$ and $A_3$ and also  $A_2$ and $A_3$ 
distaste $A_1$. (An analogy for this scenario is the case where a country has a conflict with two other countries which they are neither allied nor hostile.) The possible $\Gamma$-matrix for expressing this kind of interaction is
\begin{equation}
\Gamma_4=\left( \begin{array}{ccc}
0 & 1 & 1 \\ 
1 & 0 & 0 \\ 
1 & 0 & 0 \end{array}
\right).
\end{equation}

\subsection{Fixed point analysis \& stability}

To access to all possible states and systems’ evolving states, it is useful to find the phase space of the parameters in the dynamical equations. The phase space trajectory represents the set of states compatible with starting from any initial condition.
The dynamics equations could be represented as a vector field in the space $s_1,  s_2,  s_3$;  the points of these vector space are:
\begin{equation}\label{eq1}
\left(\dot{s_1}(s_1,s_2,s_3),\dot{s_2}(s_1,s_2,s_3)\ ,\dot{s_3}(s_1,s_2,s_3)\right) .
\end{equation}
 Each point $r\left(s_1,s_2,s_3\right)$  of space evolves in time from $t_0$ to $t_n$.
\begin{equation}\label{eq2}
r\left(t_0\right)\to r\left(t_1\right)\to r\left(t_2\right)\to \dots \to r(t_n)
\end{equation}
This evolution path draws a trajectory in the space $s_1, s_2, s_3$; each vector shows the "speed" and the evolving direction in this space. 

In order to investigate the dynamics of the system, a fixed point analysis can be fruitful. A fixed point of a function is a point which is  mapped to itself by the function. In other words,  the evolution of the equations stops at the fixed points, thus the time derivative in the evolution equations is equal to zero.  At the fixed point $ \left(s_1^*  ,s_2^*  ,s_3^*\right)$, each component of the vector field vanishes. 
The fixed points are found by solving Eqs. (\ref{eq3}):
\begin{equation}\label{eq3}
\dot{s_1}\left(s_1^*  ,s_2^*  ,s_3^*\right) =0   , \dot{s_2}\left(s_1^*  ,s_2^*  ,s_3^*\right)  =0  ,  \dot{s_3}\left(s_1^*  ,s_2^*  ,s_3^*\right)  =0
\end{equation}

A fixed point is stable if the trajectory of all points in the vicinity of the fixed point tends to evolve toward the fixed point. These points are called attractors or sinks. A fixed point is called repeller or source if all the points in the vicinity of the fixed point flow away or diverge from the fixed point; it is  called a saddle point if the vicinity of the fixed point converges to the fixed point along some directions and diverges along other directions \cite{Strogatz}.

 If the derivative of the vector field along a specified direction at the fixed point is positive, the fixed point is an attractor along this direction; if it is negative, the fixed point is a repeller along that direction.  Numerically or analytically,

\begin{equation}\label{eq4}
f\left(s_1,s_2,s_3\right)\equiv \dot{s_i}\left(s_1,s_2,s_3\right)\ ,  \ \ i=1,2,3
\end{equation}
\begin{equation}\label{eq5}
\begin{split}
f\left(s_1,s_2,s_3\right)\approx & f\left(s^*_1,  \ s^*_2,  \ s^*_3\right)+{\left(\frac{\partial f}{\partial s_1}\right)}_{s^*_1}\left(s_1-s^*_1\right)+{\left(\frac{\partial f}{\partial s_2}\right)}_{s^*_2}\left(s_2-s^*_2\right) \\
&+{\left(\frac{\partial f}{\partial s_3}\right)}_{s^*_3}\left(s_3-s^*_3\right)
\end{split}
\end{equation}
Using the first order coefficient of the expansion, the Jacobian matrix is 
\begin{equation}\label{eq6}
J\equiv \left( \begin{array}{ccc}
{\left(\frac{\partial \dot{s_1}}{\partial s_1}\right)}_{s^*_1} & {\left(\frac{\partial \dot{s_1}}{\partial s_2}\right)}_{s^*_2} & {\left(\frac{\partial \dot{s_1}}{\partial s_3}\right)}_{s^*_3} \\ 
{\left(\frac{\partial \dot{s_2}}{\partial s_1}\right)}_{s^*_1} & {\left(\frac{\partial \dot{s_2}}{\partial s_2}\right)}_{s^*_2} & {\left(\frac{\partial \dot{s_2}}{\partial s_3}\right)}_{s^*_3} \\ 
{\left(\frac{\partial \dot{s_3}}{\partial s_1}\right)}_{s^*_1} & {\left(\frac{\partial \dot{s_3}}{\partial s_2}\right)}_{s^*_2} & {\left(\frac{\partial \dot{s_3}}{\partial s_3}\right)}_{s^*_3} \end{array}
\right)
\end{equation}
One can probe the stability of a fixed point by evaluating the eigenvalues and eigenvectors of the Jacobian matrix computed at each corresponding fixed point.
\begin{equation}\label{eq7}
J\ \overrightarrow{u_i}={\lambda }_i\overrightarrow{u_i}
\end{equation}
In the above equation, $\overrightarrow{u_i}$ is  the  {\it i-}th eigenvector and ${\lambda}_i$ is the corresponding eigenvalue. 
The fixed point is an attractor along an eigenvector if the corresponding eigenvalue is negative; it is a repeller if the eigenvalue is positive;  when  all the eigenvalues are negative, the fixed point is an attractor and vice versa;  if there are  both negative and positive eigenvalues, the fixed point is a saddle point \cite{Strogatz}.

\section{RESULTS}\label{results}

The dynamics equations  are usefully rewritten as 
\begin{equation}
\dot{s_1}=s_1\left(1-s_1-s_2-s_3\right)-{\Gamma }_{12}\ s_1\ s_2\ {e}^{-{\left(\frac{s_1-s_2}{\sigma }\right)}^2}-{\Gamma }_{13}\ s_1\ s_3\ {e}^{-{\left(\frac{s_1-s_3}{\sigma }\right)}^2}
\end{equation}
\begin{equation}
\dot{s_2}=s_2\left(1-s_1-s_2-s_3\right)-{\Gamma }_{12}\ s_1\ s_2\ {e}^{-{\left(\frac{s_1-s_2}{\sigma }\right)}^2}
\end{equation}
\begin{equation}
\dot{s_3}=s_3\left(1-s_1-s_2-s_3\right)-{\Gamma }_{13}\ s_1\ s_3\ {e}^{-{\left(\frac{s_1-s_3}{\sigma }\right)}^2}.
\end{equation}

In the following, we solve the equations  for  each  $\Gamma$.

\subsection{One-way love $\Gamma_1$ type}
\begin{equation}\Gamma_1=\left( \begin{array}{ccc}
0 & -1 & -1 \\ 
1 & 0 & 0 \\ 
1 & 0 & 0 \end{array}
\right)
\end{equation}	

{\bf1.} $r_0=\left(0,0,0\right)$ is a trivial fixed point of the system. It is unstable, the eigenvalues of the Jacobian matrix are positive ${\lambda _1}=1,  \ {\lambda }_2=1,  \ {\lambda }_3=1$.

{\bf2.} $r_1=\left(1,0,0\right)$ is an attractor fixed point;  its eigenvalues are  ${\lambda _1}=-1,  \ {\lambda }_2=-1/e \equiv -0.3679,  \ {\lambda }_3=-0.3679$.   The corresponding eigenvectors are  $\overrightarrow{u_1}=(1,0,0)$ ,  $\overrightarrow{u_2}=(-1,1,0)$ and  $\overrightarrow{u_3}=(-1,0,1)$.

\begin{figure} \begin{center}
\includegraphics[width=0.5\textwidth]{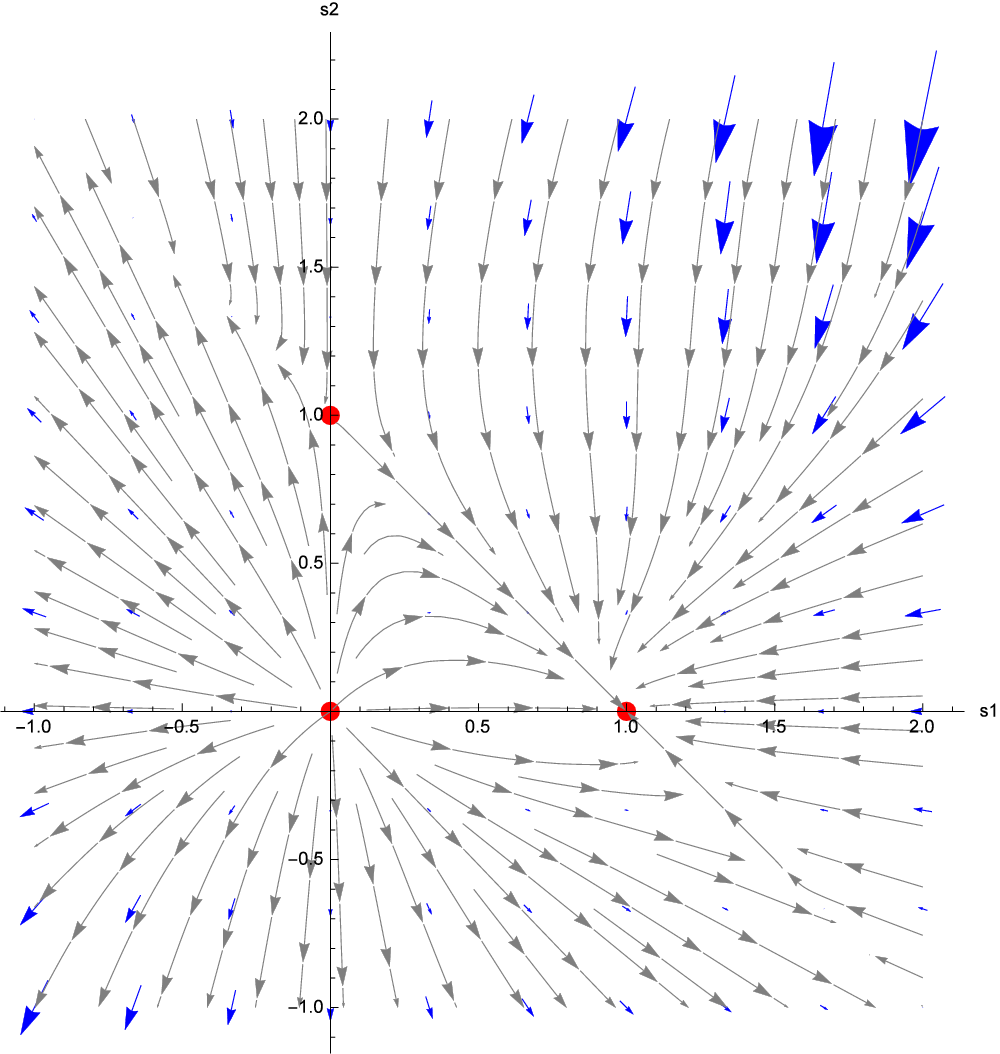}
\caption{Vector field in the plane $s_3=0$ for $\Gamma_1$. This plane contains a stable  fixed point $ r_1=(1,0,0)$ and two unstable fixed points $r_0=(0,0,0)$, $r_2=(0,1,0)$
}\label{fig: VF 2D gamma1 for s3=0}
  \end{center} \end{figure}

{\bf3.} $r_2=\left(0,1,0\right)$ is a saddle point with the eigenvalues   ${\lambda _1}=-1,  \ {\lambda }_2=0.3679,  \ {\lambda }_3=0$;  the corresponding eigenvectors are $\overrightarrow{u_1}=(0,1,0)$ , $\overrightarrow{u_2}=(1,-1,0)$ and $\overrightarrow{u_3}=(0,-1,1)$. It is an attractor along vector $\overrightarrow{u_1}=(0,1,0)$ and repeller along two other directions $u_2$ and $u_3$, so all the points on the axis $s_2$ end up to this point; see Fig. \ref{fig: VF 2D gamma1 for s3=0}. 

\begin{figure} \begin{center}
\includegraphics[width=0.5\textwidth]{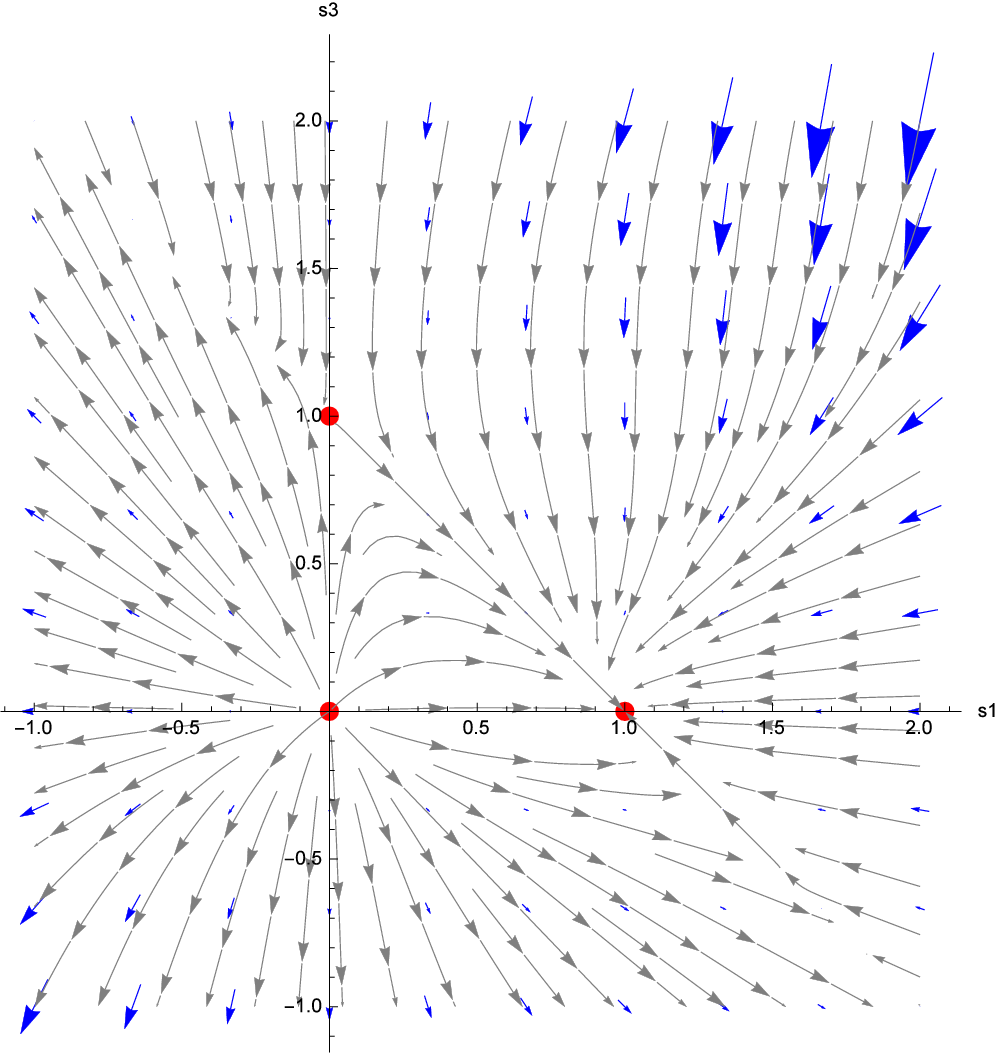}
\caption{Vector field in the plane $s_2=0$ for $\Gamma_1$. This plane contains a stable fixed point $ r_1=(1,0,0)$ and two unstable  fixed points $r_0=(0,0,0)$, $r_3=(0,0,1)$
}\label{fig: VF 2D gamma1 for s2=0}
  \end{center}\end{figure}

{\bf4.} $r_3=\left(0,0,1\right)$  is similar to the point $r_2$ due to the symmetry of the evolution equations; it is a saddle point;  its eigenvalues are ${\lambda _1}=-1,  \ {\lambda }_2=0.3679,  \ {\lambda }_3=0$; the eigenvectors are $\overrightarrow{u_1}=(0,0,1)$ , $\overrightarrow{u_2}=(1,0,-1)$ and $\overrightarrow{u_3}=(0,1,-1)$. All the points along the direction $u_1$ converge to $r_3$, whence the basin of attraction for this fixed point is the axis $s_3$; see  Fig. \ref{fig: VF 2D gamma1 for s2=0}    

{\bf5.} There is a line of fixed points in the plane $s_1=0$. All the points on the line $L=\left(0,s_2,1-s_2\right)$ are fixed point and the eigenvalues are:
\begin{equation}
{\lambda}_1=0 , {\lambda}_2=-1 , {\lambda}_3={e}^{-(1-s_2)^2-s_2^2}\ ({e}^{s_2^2}+{e}^{(1-s_2)^2}\ s_2-{e}^{s_2^2}\ s_2) 
\end{equation}

\begin{figure} \begin{center}
\includegraphics[width=0.5\textwidth]{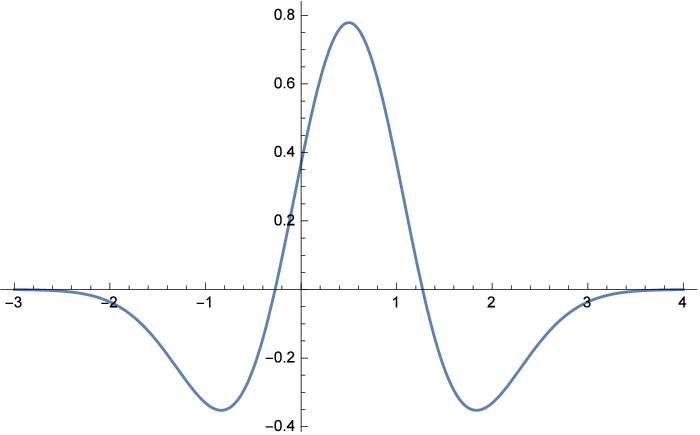}
\caption{  Eigenvalue $\lambda_3$  for the  $\Gamma_1$ case.
}\label{fig: lambda gamma1}
  \end{center}\end{figure}

\begin{figure} \begin{center}
\includegraphics[width=0.5\textwidth]{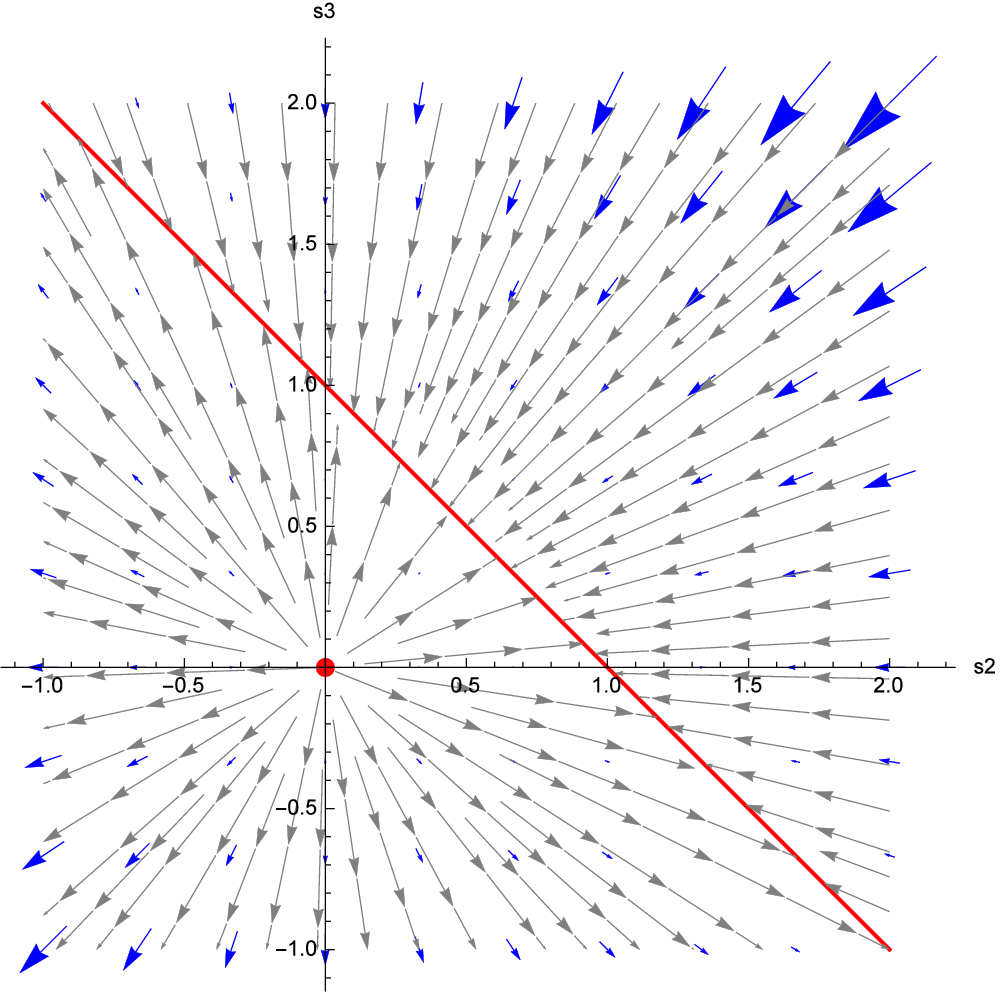}
\caption{Vector field in the plane $s_1=0$ for $\Gamma_1$
}\label{fig: VF 2D gamma1 for s1=0}
  \end{center}\end{figure}

The eigenvalue $\lambda_3$ is greater than 0 in the region $0<s_1,s_2,s_3<1$, as shown on Fig. \ref{fig: lambda gamma1}; whence these points are unstable and saddle points. The eigenvector corresponding to the eigenvalue $\lambda_2=-1$ is  $\overrightarrow{u_2}=(0,-\frac{s_2}{s_2-1},1)$ and therefore all the points in the plane $s_1=0$ converge to this line along the direction $u_2$; Fig. \ref{fig: VF 2D gamma1 for s1=0}.  

The results are summarized in   Table \ref{Table Gamma2}

\begin{table}[!h]
 \centering
 \begin{tabular}{|c|c|c|}
 \hline
{\bf Fixed Point} & {\bf Stability} & {\bf Basin of attraction } \\ \hline
$r_0=(0,0,0)$ & repeller & - \\ \hline
$r_1=(1,0,0)$ & attractor & Global \\ \hline
$r_2=(0,1,0)$ & saddle & Axis $s_2$ \\ \hline
$r_3=(0,0,1)$ & saddle & Axis $s_3$ \\ \hline
The line $L=(0,s_2,1-s_2 )$ & saddle & The surface $s_1=0$ \\ \hline
\end{tabular} \caption{ Fixed points, stability regime, basin of attraction in the $\Gamma_1$ case. } \label {Table Gamma1}
\end{table}

In this case, the vector field is perpendicular to the normal vector of the surface $s_2=s_3$; thus, the field is tangent to this surface and the trajectory  remains on this surface forever and converges to the fixed point $r_1=(1,0,0)$ for all the initial values. This means that if the two lovers $A_2$ and $A_3$ were initially and equally interested in $A_1$, their amount of   love remains equal for ever. On the other hand, if none of  $A_2$ and $A_3$ are  interested in $A_1$, their amount of initial love, decreases and converges to zero; this is in contrast to the love of $A_1$  which increases and goes to  the full capacity. When the strength of the love interaction between $A_1$ and the two lovers $A_2$, $A_3$ is equal,   no one can win the love competition. 

The point $r_1=\left(1,0,0\right)$ is the only stable fixed point;  all the points in the space except those on the plane $s_1=0$ converge to this point. This result seems intuitive. As the two lovers are not interested in $A_1$ at all, $A_1$ cannot impress them but as $A_1$ is fascinated by  both $A_2$ and $A_3$ the amount of its love increases; it will end up to be deeply in love with  both $A_2$ and $A_3$. 

\begin{figure} \begin{center}
\includegraphics[width=0.5\textwidth]{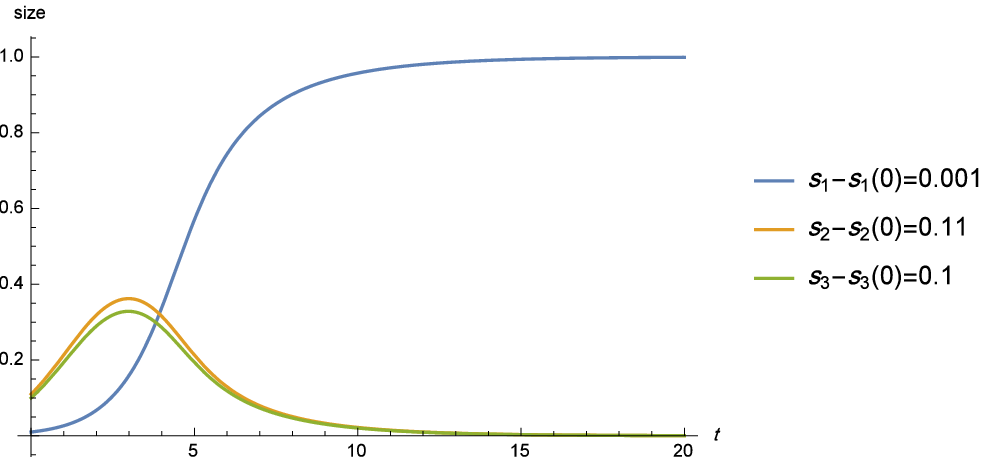}
\caption{Time evolution of the size of love  for the  case $\Gamma_1$
}\label{fig: Time Evolution gamma1} 
  \end{center} \end{figure}

\begin{figure} \begin{center}
\includegraphics[width=0.5\textwidth]{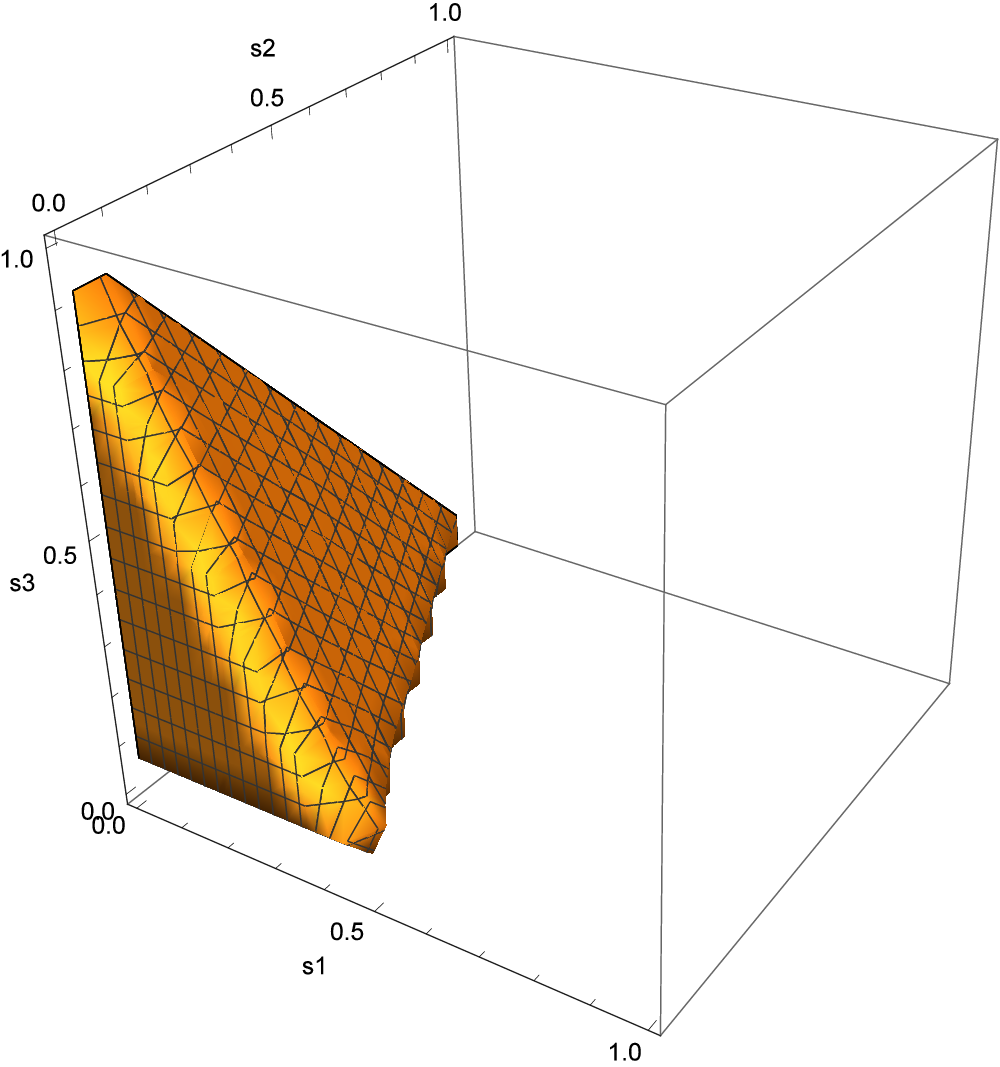}
\caption{ Phase space region of weak attraction between the 3  lovers for the $\Gamma_1$ case.
}\label{fig: Region gamma1}
  \end{center} \end{figure}

We also find an interesting situation when the lover $A_1$ starts with a little love at the beginning of the game and when the lovers $A_2$ and $A_3$ get somehow interested in $A_1$; both  demand more attention from the lover $A_1$; therefore their love increases at  first but later they become less interested and their love decreases,  ending up to zero; see Fig. \ref{fig: Time Evolution gamma1}.
This case  occurs when the initial $s_1$ is small and for $1-s_1-s_2-s_3>0$; in such a case,  $\dot{s_2}>0$ and $\dot{s_3}>0$. One can numerically find a region in which $s_2$ and $s_3$ increase at first; see  Fig.   \ref{fig: Region gamma1}.  

On the surface $s_1=0$, there is a line of fixed points;  the basin of attraction is the surface $s_1=0$; this means that if the lover $A_1$ wishes neither $A_2$ nor $A_3$ at first it will remain uninterested, but there will be a competition between $A_2$ and $A_3$ to obtain $A_1$'s attention;  the initial ratio of their love $\frac{s_2}{s_3}$ remains unchanged and their interest reaches a balance state on the line $L=\left(0,s_2,1-s_2\right)$; see Fig. \ref{fig: VF 2D gamma1 for s1=0} again. 

\subsection{One-way love $\Gamma_2$ type}
\begin{equation}\Gamma_2=\left( \begin{array}{ccc}
0 & 1 & 1 \\ 
-1 & 0 & 0 \\ 
-1 & 0 & 0 \end{array}
\right)
\end{equation}	

The fixed points, stability regimes, and basins of attraction  are summarized in Table \ref{Table Gamma2}.

{\bf1.}  $r_0=\left(0,0,0\right)$ is an unstable fixed point and its eigenvalues are ${\lambda _1}=1,  \ {\lambda }_2=1,  \ {\lambda }_3=1$; 

{\bf2.} $r_1=\left(1,0,0\right)$ is a saddle point with eigenvalues ${\lambda _1}=-1,  \ {\lambda }_2=0.3679,  \ {\lambda }_3=-0.3679$.  The eigenvector corresponding to the eigenvalue ${\lambda_1}=-1$ is  $\overrightarrow{u_1}=(1,0,0)$; thus, it is an attractor along the axis $s_1$; 

\begin{figure} \begin{center}
\includegraphics[width=0.5\textwidth]{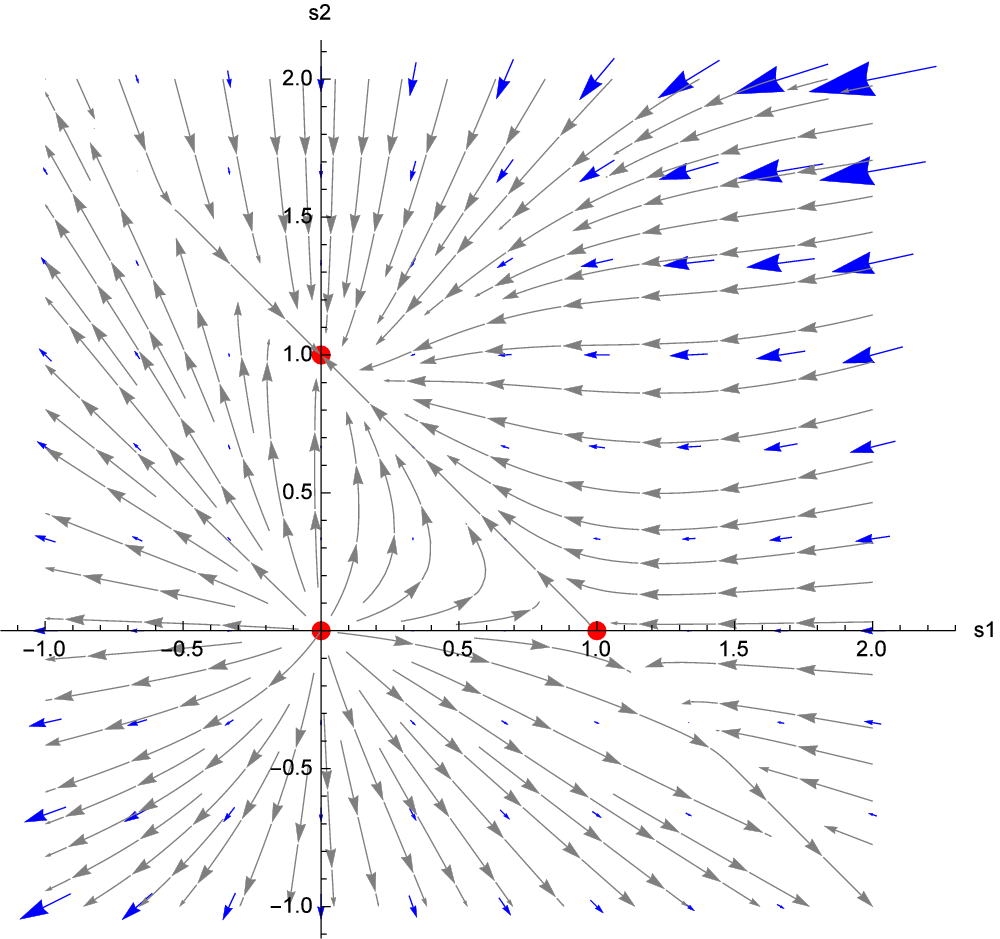}
\caption{Vector field in the plane $s_3=0$ for $\Gamma_2$.This plane contains a stable point $ r_2=(0,1,0)$ and two unstable points $r_0=(0,0,0)$, $r_1=(1,0,0)$
}\label{fig: VF 2D gamma2 for s3=0}
  \end{center}\end{figure}

{\bf3.} $r_2=\left(0,1,0\right)$ is a stable fixed point: the eigenvalues are ${\lambda _1}=-1,  \ {\lambda }_2=-0.3679,  \ {\lambda }_3=0$;  the corresponding eigenvectors are   $\overrightarrow{u_1}=(0,1,0)$ , $\overrightarrow{u_2}=(-1,1,0)$ and $\overrightarrow{u_3}=(0,-1,1)$;  see   Fig. \ref{fig: VF 2D gamma2 for s3=0}; 

{\bf4.} $r_3=\left(0,0,1\right)$ is similar to the point $r_2$; thus, it is a stable fixed point; he eigenvalues are ${\lambda _1}=-1,  \ {\lambda }_2=-0.3679,  \ {\lambda }_3=0$; the corresponding eigenvectors are   $\overrightarrow{u_1}=(0,0,1)$ , $\overrightarrow{u_2}=(1,0,-1)$ and $\overrightarrow{u_3}=(0,1,-1)$;

{\bf5.} There is a line of fixed points in the plane $s_1=0$. All the points on the line $L=\left(0,s_2,1-s_2\right)$ are fixed points;  their eigenvalues are:
\begin{equation}
{\lambda}_1=0 , {\lambda}_2=-1 , {\lambda}_3={e}^{-(1-s_2)^2-s_2^2}\ (-{e}^{s_2^2}-{e}^{(1-s_2)^2}\ s_2+{e}^{s_2^2}\ s_2) 
\end{equation}

\begin{figure} \begin{center}
\includegraphics[width=0.5\textwidth]{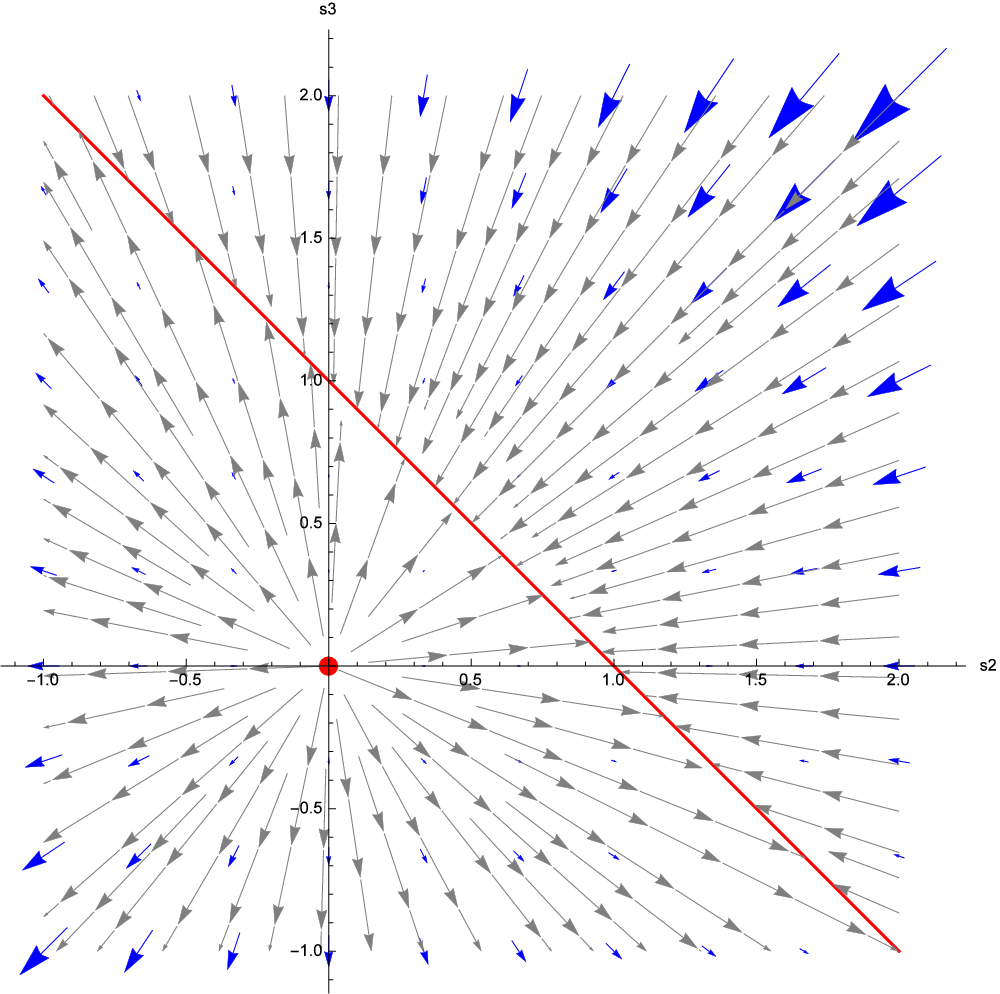}
\caption{Vector field in the plane $s_1=0$ for the $\Gamma_2$ case.
}\label{fig: VF 2D gamma2 for s1=0}
  \end{center}\end{figure}

The eigenvalue $\lambda_3$ is  negative in the region $0<s_1,s_2,s_3<1$, so that  these points are stable and attractors. The eigenvector corresponding to the eigenvalue $\lambda_2=-1$ is  $\overrightarrow{u_2}=(0,-\frac{s_2}{s_2-1},1)$;  therefore all the points in the plane $s_1=0$ converge to this line along the direction $u_2$; see Fig.  \ref{fig: VF 2D gamma2 for s1=0}. 

\begin{table}[!h]
 \centering
 \begin{tabular}{|c|c|c|}
 \hline
{\bf Fixed Point} & {\bf Stability} & {\bf Basin of attraction} \\ \hline
$r_0=(0,0,0)$ & repeller & - \\ \hline
$r_1=(1,0,0)$ & saddle & Axis $s_2$ \\ \hline
$r_2=(0,1,0)$ & stable & The surface $s_3=0$ \\ \hline
$r_3=(0,0,1)$ & stable& The surface $s_2=0$ \\ \hline
The line $L=(0,s_2,1-s_2 )$ & stable& Global \\ \hline
\end{tabular} \caption{ Fixed points, stability regime, basin of attraction in the $\Gamma_2$ case. } \label {Table Gamma2}
\end{table}

All the points in the space except the axis $s_1$ converge to this line. This result seems intuitive, as $A_1$ is not interested in $A_2$ and $A_3$ at all, $A_1$ cannot be impressed by the lovers  $A_2$ and $A_3$. But as $A_2$ and $A_3$ are fascinated by $A_1$ the amount of their love increases up to be deeply  in love with $A_1$ forever.

\begin{figure} \begin{center}
\includegraphics[width=0.5\textwidth]{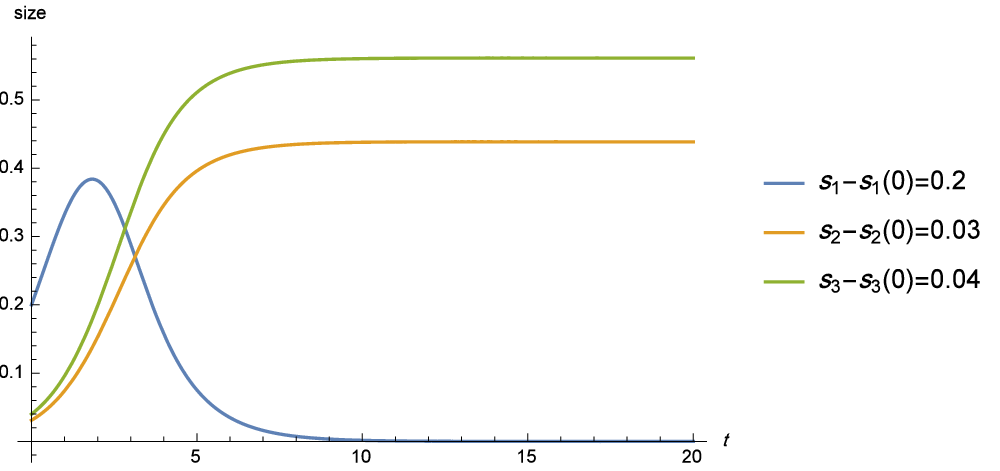}
\caption{Example of  time evolution of the size of love for the case $\Gamma_2$
}\label{fig: Time Evolution gamma2}
\end{center}\end{figure}

\begin{figure} \begin{center}
\includegraphics[width=0.5\textwidth]{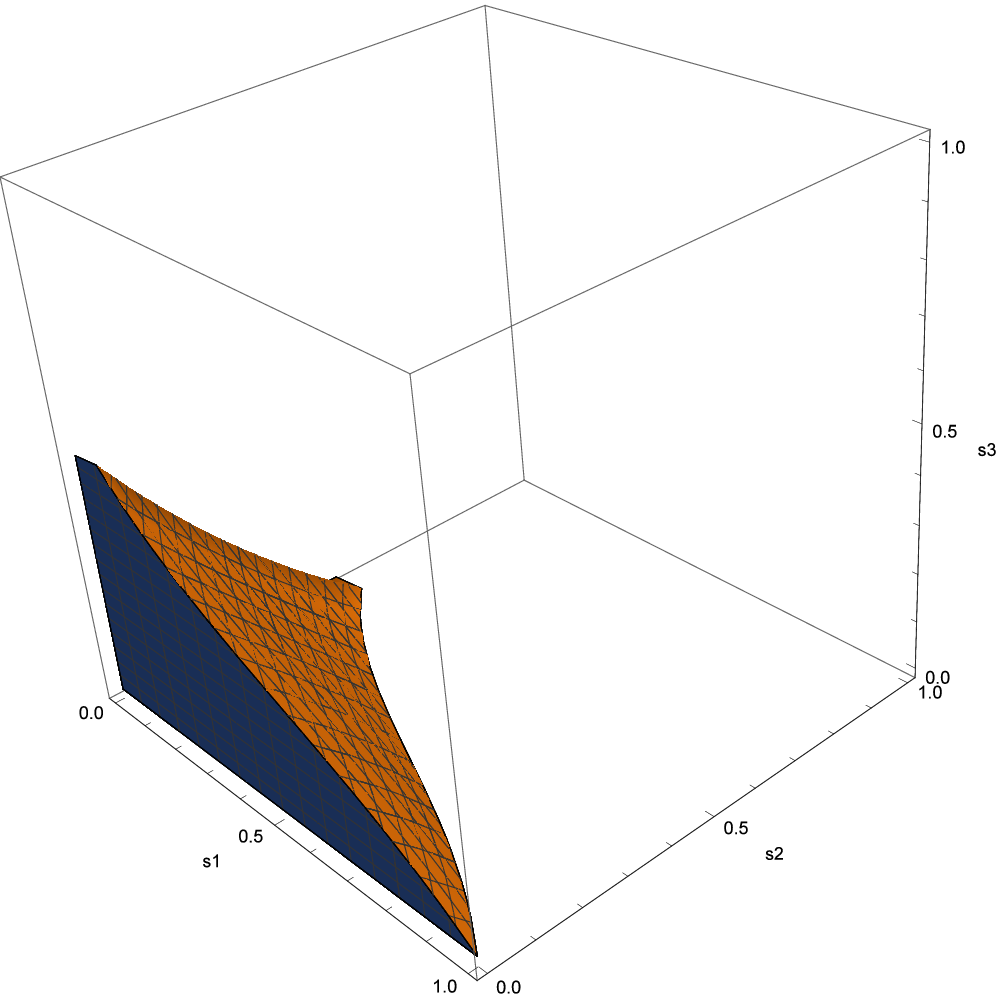}
\caption{ Numerical calculation and display of the  region in which $s_1$  increases at first, in the case $\Gamma_2$.
}\label{fig: Region gamma2}
\end{center}\end{figure}

When the initial love of the lovers $A_2$ and $A_3$ is small and  under the condition $1-s_1-s_2-s_3>0$ with  $\dot{s_1}>0$, the lover $A_1$ gets somehow interested in  $A_2$ and $A_3$ and demands more attention from  these lovers;  whence  its love increases at   first. As an example consider a movie star at the beginning of its work when he or she is not a celebrated person and is known by a few people, the movie star is interested in  getting fans and   greedily demands their attention. But later on,  $A_1$ looses  to  be interested; thereafter $s_1$  decreases and ends up at zero;  see Fig. \ref{fig: Time Evolution gamma2}.
One can numerically find a region in which $s_1$  increases at first; Fig. \ref{fig: Region gamma2}.

Just like in  the previous section, one finds  the vector field  to be tangent to the surface $s_2=s_3$; all the trajectories remain on this surface forever and converge to the fixed point $r_1=(0,0.5,0.5)$.

\subsection{Love triangle $\Gamma_3$ type}
\begin{equation}\Gamma_3=\left( \begin{array}{ccc}
0 & -1 & -1 \\ 
-1 & 0 & 0 \\ 
-1 & 0 & 0 \end{array}
\right)
\end{equation}

\begin{figure} \begin{center}
\includegraphics[width=0.5\textwidth]{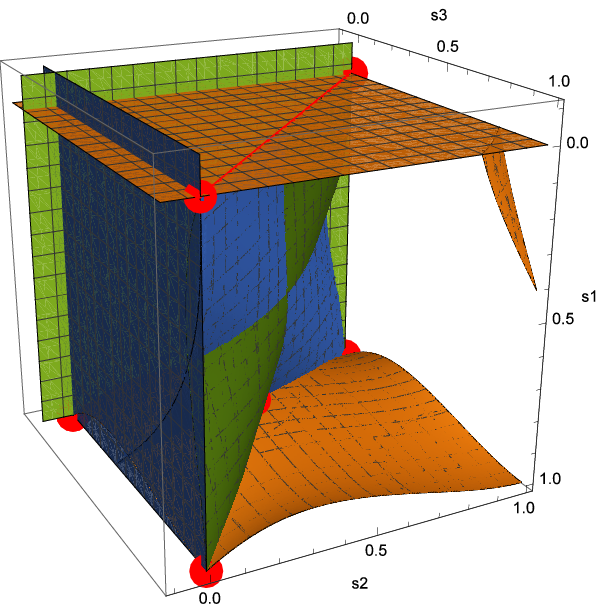}
\caption{For the  $\Gamma_3$ case, the three  evolution surfaces $\dot{s_1} =0$, $\dot{s_2} =0$ and $\dot{s_3} =0$  can be displayed in the phase space as:  the blue   surface  corresponds to $\dot{s_1} =0$, the green one is   for $\dot{s_2} =0$, and the  $\dot{s_3} =0$  surface is illustrated in orange. The fixed points are indicated by  the red points.}\label{fig: contour plot 3D for gamma3}
  \end{center}\end{figure}

{\bf1.} $r_0=\left(0,0,0\right)$ is a trivial fixed point of the system.
It is unstable, the eigenvalues of the Jacobian matrix are positive ${\lambda _1}=1,  \ {\lambda }_2=1,  \ {\lambda }_3=1$.

{\bf2.} There are three fixed points on the surface $s_2=0$. $ r_1=(1,0,1)$ , $r_2=(1,0,0)$, $r_3=(0,0,1)$;
$r_1$ is an attractor fixed point;  its eigenvalues are ${\lambda _1}=-1,  \ {\lambda }_2=-1,  \ {\lambda }_3=  -1+1/e \equiv    -0.632 $. 

\begin{figure} \begin{center}
\includegraphics[width=0.5\textwidth]{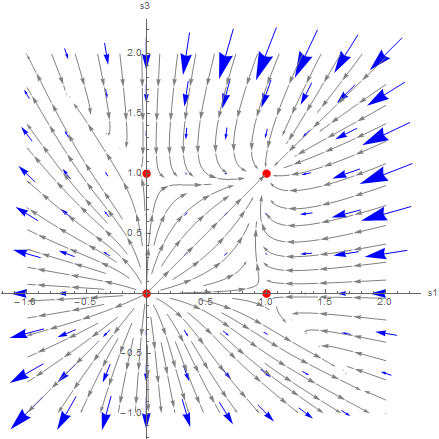}
\caption{The vector field in the plane $s_2=0$ for $\Gamma_3$. This plane contains a stable fixed point $ r_1=(1,0,1)$  and three unstable fixed points $r_0=(0,0,0)$, $r_2=(1,0,0)$, $r_3=(0,0,1)$.
}\label{fig: VF 2D gamma3 for s2=0}
  \end{center}\end{figure}

The eigenvalues of the point $r_2$ are ${\lambda _1}=-1,  \ {\lambda }_2=0.3679,  \ {\lambda }_3=0.3679$; thus it is  a saddle point. The eigenvector corresponding to ${\lambda}_1=-1$ is $\overrightarrow{u_1}=(1,0,0)$. Therefore, it is an attractor along the axis $s_1$. On the other hand, it is a repeller along the two directions $\overrightarrow{u_1}=(-0.42 ,0.91 ,0)$ and $\overrightarrow{u_1}=(-0.42 ,0 ,0.91)$;  Fig. \ref{fig: VF 2D gamma3 for s2=0}.

Similarly, the point $r_3$ is a saddle point with the eigenvalues  ${\lambda _1}=-1,  \ {\lambda }_2=0.3679,  \ {\lambda }_3=0$.  The eigenvector corresponding to ${\lambda}_1=-1$ is $\overrightarrow{u_1}=(0,0,1)$. Therefore, it is an attractor along the axis $s_3$.

{\bf3.} There are two other fixed points $r_4=(1,1,0)$, $r_5=(0,1,0)$ on the $s_3=0$ plane.

The fixed point $r_4$ is similar to the fixed point $r_1$ and its eigenvalues are the same ${\lambda _1}=-1,  \ {\lambda }_2=-1,  \ {\lambda }_3=-0.632$;  thus, it is an attractor fixed point.

\begin{figure} \begin{center}
\includegraphics[width=0.5\textwidth]{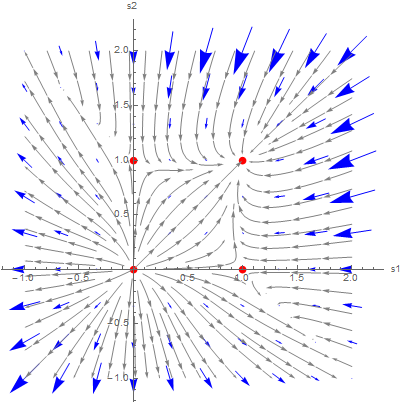}
\caption{Vector field in the plane $s_3=0$ for $\Gamma_3$, it is similar to the vector field in the plane $s_2=0$ due to the symmetry in dynamical equations.This plane contains a stable  fixed point $ r_4=(1,1,0)$ and three unstable fixed points $r_0=(0,0,0)$, $r_2=(1,0,0)$, $r_5=(0,1,0)$.
}\label{fig: VF 2D gamma3 for s3=0} 
  \end{center}\end{figure}

The eigenvalues of the point $r_5$ are ${\lambda _1}=-1, \ {\lambda }_2=0.3679, \ {\lambda }_3=0$. The eigenvector corresponding to ${\lambda_1=-1}$ is $\overrightarrow{u_1}=(0,1,0)$. So this point is a saddle point which attracts along the axis $s_2$; see Fig. \ref{fig: VF 2D gamma3 for s3=0}. 

{\bf4.} There exists a line of fixed points in the plane $s_1=0$. In other words, all the points on the line $L=(0,s_2,1-s_2 )$ are fixed points. These points are saddle points and their eigenvalues are:  
\begin{equation}
{\lambda}_1=0 , {\lambda}_2=-1 , {\lambda}_3={e}^{-(1-s_2)^2-s_2^2}\ ({e}^{s_2^2}+{e}^{(1-s_2)^2}\ s_2-{e}^{s_2^2}\ s_2) 
\end{equation}

\begin{figure} \begin{center}
\includegraphics[width=0.5\textwidth]{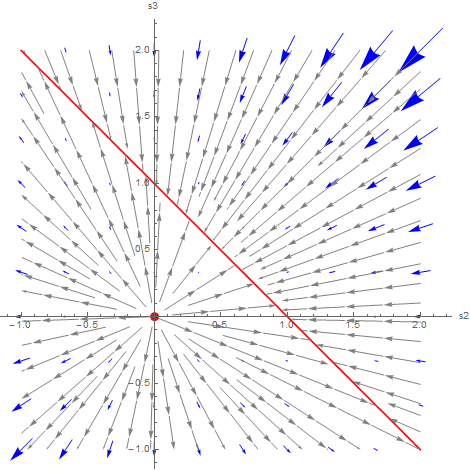}
\caption{Vector field in the plane $s_1=0$ for the $\Gamma_3$ case. }\label{fig: VF 2D gamma3 for s1=0}
  \end{center}\end{figure}

Since  $ 0<s_1,s_2,s_3<1$, the eigenvalue ${\lambda}_3$ is greater than 0 in this interval, so these points are saddle points. The eigenvector corresponding to ${\lambda}_2=-1$ is $\overrightarrow{u_2}=(0 ,-\frac{s_2}{-1+s_1}  ,1)$ and is located in the plane $s_1=0$; thus,  all the points in this plane converge to the line of fixed points;  see Fig. \ref{fig: VF 2D gamma3 for s1=0}. 

{\bf5.}  The  $r_6$  fixed point  $\approx (0.8565, 0.4282, 0.4282)$  is an unstable saddle point. Indeed, the eigenvalues are ${\lambda}_1\approx-1.2615  ,{\lambda}_2\approx-0.7130 ,{\lambda}_3\approx0.2615$;  the corresponding eigenvectors are:
\begin{equation}
\overrightarrow{u_1}=(-0.8165 ,-0.4082 ,-0.4082)
\end{equation}
\begin{equation}
\overrightarrow{u_2}=(0.2429 ,0.6859 ,0.6859)
\end{equation}
\begin{equation}
\overrightarrow{u_3}=(0 ,-0.7071 ,0.7071)
\end{equation}

\begin{figure} \begin{center}
\includegraphics[width=0.5\textwidth]{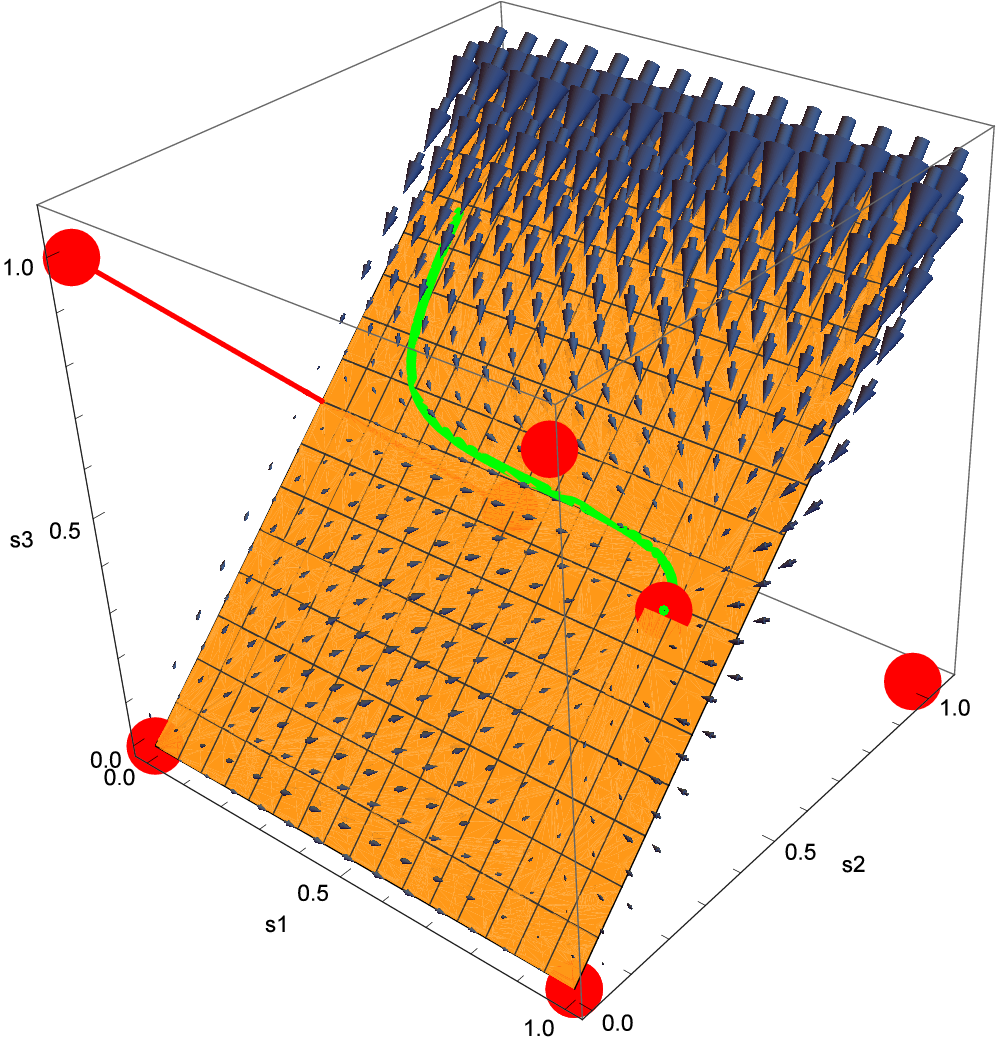}
\caption{Vector field in the plane $s_2=s_3$ for $\Gamma_3$; fixed points indicated by red dots.}\label{fig: VF 3D gamma3 no.2}
  \end{center}\end{figure}

This point is a repeller along the third eigenvector $\overrightarrow{u_3}$ with eigenvalue ${\lambda}_1\approx-1.2615$.
Thus,  all the trajectories perpendicular to this vector at this point will  not be repelled. The surface $s_2=s_3$ is perpendicular to the vector $\overrightarrow{u_3}$; the inner product of vector field $\overrightarrow{\dot{S}}=(\dot{s_1},\dot{s_1},\dot{s_1})$ lies on the surface $s_2=s_3$; the normal vector to the surface $\overrightarrow{u_3}$ is zero:
\begin{equation}
\overrightarrow{\dot{S}}\bullet \overrightarrow{u_3}=0
\end{equation}
Therefore, all the trajectories on the surface $s_2=s_3$ remain on this surface and converge to the fixed point $r_6$. Thus,  the surface $s_2=s_3$ is the basin of attraction of this point and all the trajectories on this surface converge to the fixed point $r_6$; Fig\ref{fig: VF 3D gamma3 no.2}.

\begin{figure} \begin{center}
\includegraphics[width=0.5\textwidth]{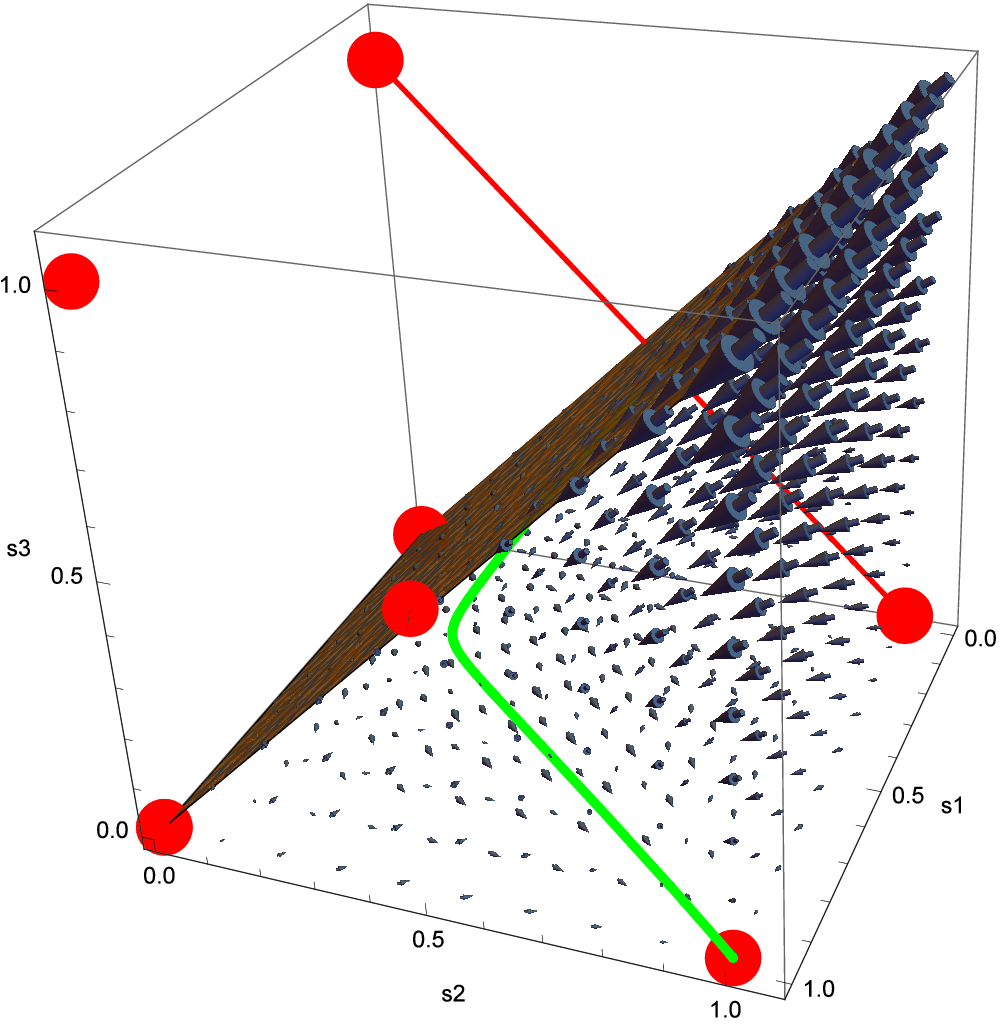}
\caption{  The vector field in the region $s_3<s_2$ for $\Gamma_3$. The trajectories  of an arbitrary initial point in the region $s_3<s_2$ converge to $r_4=(1,1,0)$; fixed points indicated by red dots.}\label{fig: VF 3D gamma3 no.1} 
  \end{center}\end{figure}

\begin{figure} \begin{center}
\includegraphics[width=0.5\textwidth]{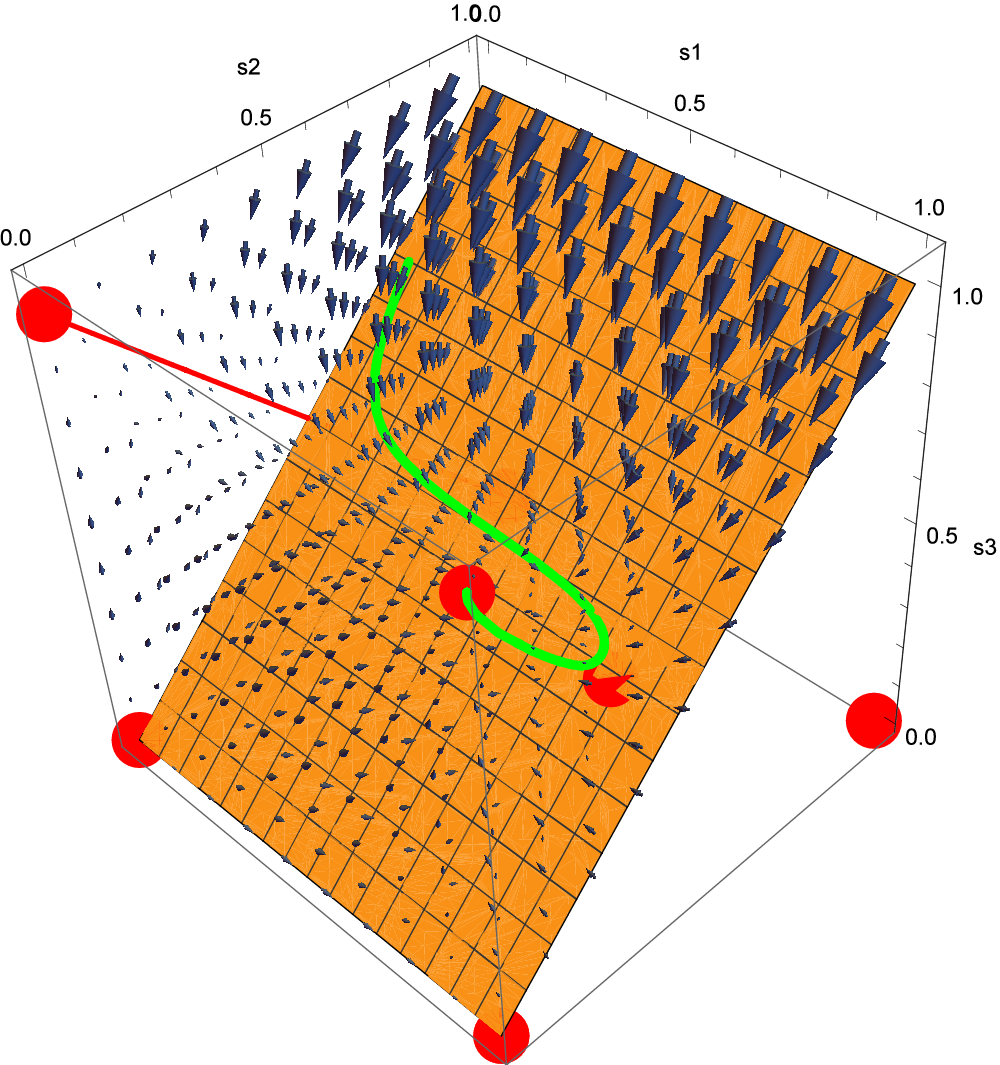}
\caption{The vector field in the region $s_3>s_2$ for $\Gamma_3$. The trajectory of an arbitrary initial point in the region $s_3>s_2$ converges to $r_1=(1,0,1)$.}\label{fig: VF 3D gamma3 no.3}
  \end{center}\end{figure}

The  trajectories which  are not on the surface  $s_2=s_3$ diverge from the surface. The points that are located under this surface  $s_3<s_2$ converge to the stable fixed point $r_4=(1,1,0)$ (Fig. \ref{fig: VF 3D gamma3 no.1});  the points that are located above this surface $s_3>s_2$ converge to the fixed point $r_1=(1,0,1)$; see Fig. \ref{fig: VF 3D gamma3 no.3}.

The results are summarized  in Table \ref{Table Gamma3}.

So,  two fixed points  $r_1=\left(1,0,1\right)$ and $r_4=\left(1,1,0\right)$ correspond to a situation in which one of the lovers wins the love game in a love triangle and can get all the attention of the lover $A_1$;  if one of the lovers $A_2$ and $A_3$ is initially more interested in $A_1$, it will capture  the full attention of the lover $A_1$.

But if the two lovers love $A_1$ equally at first, no one wins the game and both share the love and attention of $A_1$, -  and reach   the point $r_6$. As this fixed point is strongly unstable and if the condition changes a little bit, the love path diverges from the point $r_6$. For example, if the love of $A_2$ regresses and gets lower than $A_3$, then $A_2$ will loose the game and the love path ends up in  $r_1=\left(1,0,1\right)$.

\begin{table}[!h]
 \centering
 \begin{tabular}{|c|c|c|}
 \hline
{\bf Fixed Point} & {\bf Stability} & {\bf Basin of attraction} \\ \hline
$r_0=(0,0,0)$ & repeller & - \\ \hline
$r_1=(1,0,1)$ & attractor & $s_3>s_2$ \\ \hline
$r_2=(1,0,0)$ & saddle & Axis $s_1$ \\ \hline
$r_3=(0,0,1)$ & saddle & Axis $s_3$ \\ \hline
$r_4=(1,1,0)$ & attractor & $s_3<s_2$ \\ \hline
$r_5=(0,1,0)$ & saddle & Axis $s_2$ \\ \hline
$r_6\approx(0.8565 ,0.4282 ,0.4282)$ & saddle & The surface $s_2=s_3$ \\ \hline
The line $L=(0,s_2,1-s_2 )$ & saddle & The surface $s_1=0$ \\ \hline
\end{tabular} \caption{ Fixed points, stability regimes, basins of attraction in the $\Gamma_3$ case. } \label {Table Gamma3}
\end{table} 

\subsection{Love triangle $\Gamma_4$ type}
Now consider: \begin{equation}\Gamma_4=\left( \begin{array}{ccc}
0 & 1 & 1 \\ 
1 & 0 & 0 \\ 
1 & 0 & 0 \end{array}
\right)
\end{equation}

\begin{figure} \begin{center}
\includegraphics[width=0.5\textwidth]{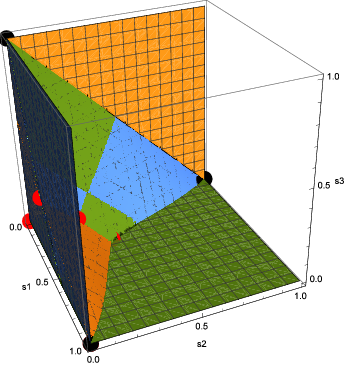}
\caption{The three  evolution surfaces $\dot{s_1} =0$, $\dot{s_2} =0$ and $\dot{s_3} =0$ for  the $\Gamma_4$ case.. The orange one is the surface  $\dot{s_1 }=0$, the blue one is $\dot{s_2}=0$ and the surface $\dot{s_3}=0$ is illustrated in green. The stable fixed points are  the  the black points and the unstable fixed points are   the red points.}\label{fig: contour plot 3D for gamma4}
  \end{center}\end{figure} 

In this case the fixed points and their stability  quite differ from the previous case. The three surfaces $\dot{s_1 }=0$, $\dot{s_2 }=0$ and $\dot{s_3}=0$ are illustrated in Fig. \ref{fig: contour plot 3D for gamma4}. 

{\bf1.} The point $r_0=(0,0,0)$ is an unstable fixed point with eigenvalues ${\lambda _1}=1,  \ {\lambda }_2=1,  \ {\lambda }_3=1$.

\begin{figure} \begin{center}
\includegraphics[width=0.5\textwidth]{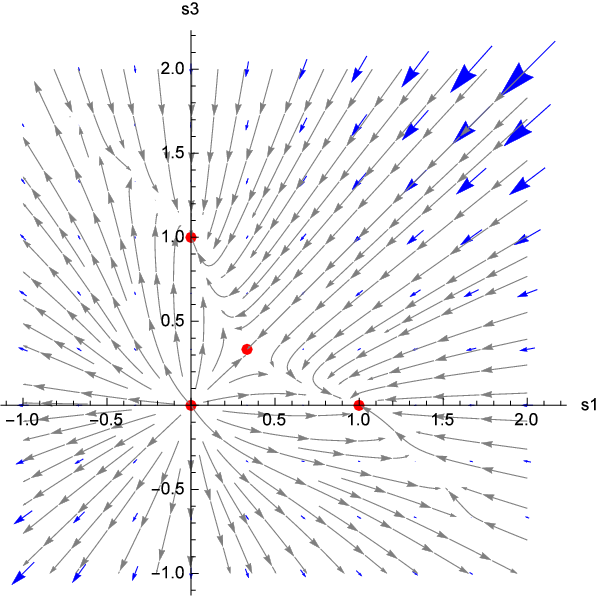}
\caption{Vector field on the plane $s_2=0$, for the $\Gamma_4$ case.}\label{fig: VF 2D gamma4 for s2=0} 
    \end{center}  \end{figure}

{\bf2.} There are three fixed points on the plane $s_2=0$.
The point $r_1=(0.3333,0 ,0.3333)$ with the eigenvalues ${\lambda _1}=-1,  \ {\lambda }_2=0.3333,  \ {\lambda }_3=0.3505$ is an unstable saddle point. It is attractor along $\overrightarrow{u_1}=(1,0,1)$ with the eigenvalue ${\lambda}_1=-1$, as shown in  Fig. \ref{fig: VF 2D gamma4 for s2=0}. 
The point $r_2=(0,0,1)$ is an attractor fixed point and its eigenvalues are ${\lambda _1}=-1,  \ {\lambda }_2=-0.3679,  \ {\lambda }_3=1-0$.
The point $r_3=(1,0 ,0)$ with eigenvalues ${\lambda _1}=-1,  \ {\lambda }_2=-0.3679,  \ {\lambda }_3=-0.3679$ is an attractor.

\begin{figure} \begin{center}
\includegraphics[width=0.5\textwidth]{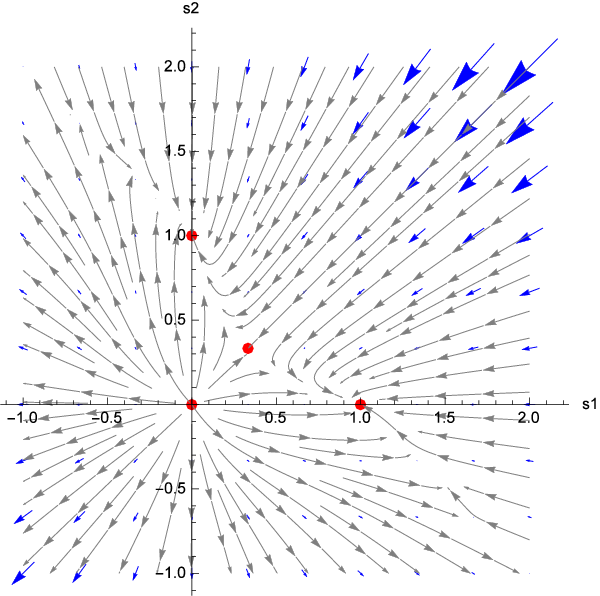}
\caption{Vector field on the plane $s_3=0$ for the $/Gamma_4$ case.}\label{fig: VF 2D gamma4 for s3=0}
  \end{center}\end{figure}

{\bf3.} There are two other fixed points on the plane $s_3=0$.
The point $r_4=(0.3333,0.3333 ,0)$ with  eigenvalues ${\lambda _1}=-1,  \ {\lambda }_2=0.3333,  \ {\lambda }_3=0.3505$ is an unstable saddle point. All the points along the eigenvector $\overrightarrow{u_1}=(1,1,0)$ converge to this point; see Fig. \ref{fig: VF 2D gamma4 for s3=0}.  
The point $r_5=(0,1 ,0)$ with eigenvalues ${\lambda _1}=-1,  \ {\lambda }_2=-0.3679,  \ {\lambda }_3=0$ is an attractor.

\begin{figure} \begin{center}
\includegraphics[width=0.5\textwidth]{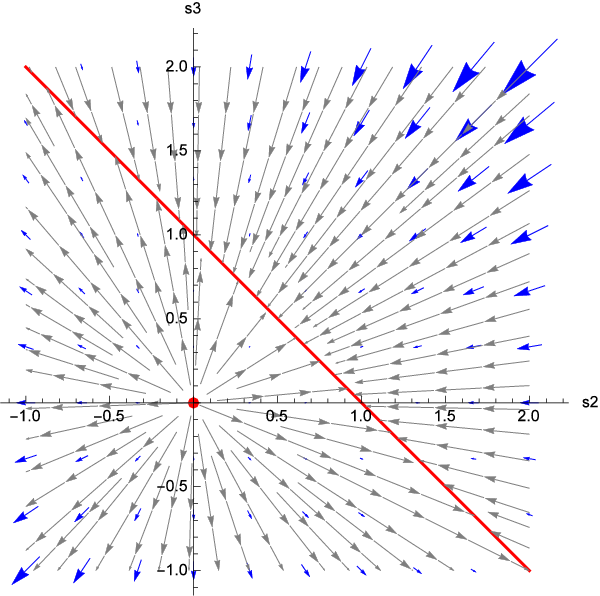}
\caption{Vector field in the plane $s_1=0$, for the $\Gamma_4$ case. }\label{fig: VF 2D gamma4 for s1=0}
  \end{center}\end{figure}

{\bf4.} There is a line of fixed points in the plane $s_1=0$. All the points on the line $L=(0,s_2,1-s_2 )$ are fixed points with eigenvalues:
\begin{equation}
{\lambda}_1=0 , {\lambda}_2=-1 , {\lambda}_3={e}^{-(1-s_2)^2-s_2^2}\ (-{e}^{s_2^2}-{e}^{(1-s_2)^2}s_2+{e}^{s_2^2}s_2) 
\end{equation}
The third eigenvector ${\lambda}_3$ is negative in the region $0<s_2<1$. So, unlike in  the previous case, all the fixed points on this line are stable: Fig.  \ref{fig: VF 2D gamma4 for s1=0}.

{\bf5.} In the region $0<s_1,s_2,s_3<1$,  there is a fixed point $r_6\approx(0.3365 ,0.1682 ,0.1682)$ which is an unstable saddle point;  its eigenvalues are ${\lambda _1}\approx-0.9815,  \ {\lambda }_2\approx0.3271,  \ {\lambda }_3\approx-0.0185$.
Consider the surface $-s_1+s_2+s_3=0$. The normal vector of this surface is $\overrightarrow{n}=(-1,1,1)$
The vector field on this surface is:

\begin{equation}
\dot{s_1}=\left(s_2+s_3\right)\ \left(1+\left(-2-{e}^{-s_3^2}\right)s_2+\left(-2-{e}^{-s_2^2}\right)s_3\right)
\end{equation}
\begin{equation}
\dot{s_2}=s_2\ \left(1-2\ s_2-2\ s_3-{e}^{-s_3^2}\left(s_2+s_3\right)\right)
\end{equation}
\begin{equation}
\dot{s_3}=s_3\ \left(1-2\ s_2-2\ s_3-{e}^{-s_2^2}\left(s_2+s_3\right)\right).
\end{equation}

\begin{figure} \begin{center}
\includegraphics[width=0.5\textwidth]{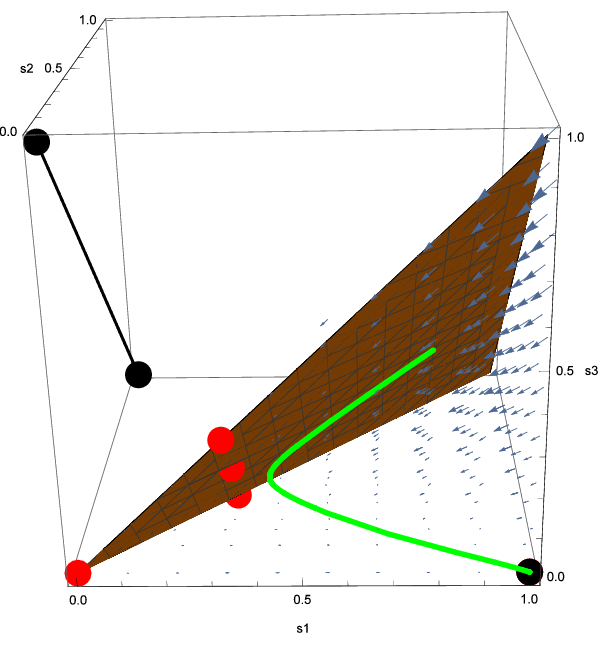}
\caption{ Illustration of the  phase space in the  $\Gamma_4$ case, when  trajectories initially located under the   $s_2+s_3<s_1$   surface  converge to the stable fixed point $r_3=(1,0 ,0)$; the opposite case is shown in Fig.    \ref{fig: VF 3D gamma4  no.2}.}\label{fig: VF 3D gamma4 no.1}
  \end{center}\end{figure}

The inner product of vector field $\overrightarrow{\dot{S}}=(\dot{s_1},\dot{s_1},\dot{s_1})$ on the surface $-s_1+s_2+s_3=0$ and the normal vector  $ \overrightarrow{n} $   to the surface,
 $\overrightarrow{n}=(-1,1,1)$,  is zero.
\begin{equation}
\overrightarrow{\dot{S}}\bullet \overrightarrow{n}=0
\end{equation}
Therefore, the vector field is perpendicular to  $ \overrightarrow{n}$; thus,   the vector field is tangent to the surface and all the trajectories  remain on this surface. These trajectories converge to $r_6\approx(0.3365 ,0.1682 ,0.1682)$. 
The  trajectories initially located under the   $s_2+s_3<s_1$  surface converge to the stable fixed point $r_3=(1,0 ,0)$ (see Fig. \ref{fig: VF 3D gamma4 no.1}); those  located above the $s_2+s_3>s_1$  surface  converge to the line $L=(0,s_2,1-s_2 )$;  see Fig.   \ref{fig: VF 3D gamma4  no.2}. 

  As an example for this case of interaction, consider three countries $A_1$, $A_2$ and $A_3$. There is a hostility between the country $A_1$ and the two countries $A_2$ and $A_3$ but there is no any hostility or alliance between the countries $A_2$ and $A_3$. 

The point  $r_3=(1,0,0)$ corresponds to a situation in which $A_1$ is interested in the two others and demands a good relationship, but $A_2$ and $A_3$ are not interested in $A_1$ and tend to boycott it.

As mentioned before there is a fixed point on the surface $s_1=s+2+s_3$  for which the basin of attraction is $s_1=s_2+s_3$. This means that under this special condition,  there is an unusual fixed point in which all the countries/lovers are interested in each other and seek  a mutually peaceful relationship, like a  "menage a 3".  However,  this balanced state is extremely unstable; any small deviation (perturbation)  from the surface, gives rise to a divergent trajectory from the fixed point, either leading to the fixed point $r_3$ or to the  line $L$.

(A "symmetric solution" is possible: the line  $L=(0,s_2,1-s_2 )$  turns  out to be  obviously the opposite of this previously  discussed situation.)

\begin{figure} \begin{center}
\includegraphics[width=0.5\textwidth]{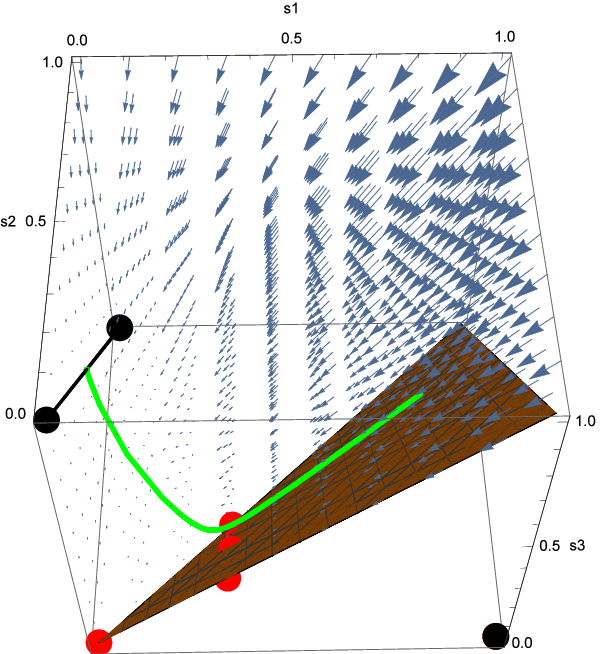}
\caption{ Illustration of the  phase space in the  $\Gamma_4$ case,  when  trajectories  initially 
located above the  $s_2+s_3>s_1$ surface  converge to the line $L=(0,s_2,1-s_2 )$; the opposite case is shown in Fig.    \ref{fig: VF 3D gamma4  no.1}}\label{fig: VF 3D gamma4 no.2}
  \end{center} \end{figure} 

The results are summarized  in  Table \ref{Table Gamma4}.
\begin{table}[!h]
 \centering
 \begin{tabular}{|c|c|c|}
 \hline
{\bf Fixed Point} & {\bf Stability} & {\bf Basin of attraction} \\ \hline
$r_0=(0,0,0)$ & repeller & - \\ \hline
$r_1=(0.3333,0 ,0.3333)$ & saddle & $s_2=0,s_1=s_3$ \\ \hline
$r_2=(0,0,1)$ & attractor & $s_2=0 ,s_1<s_3$ \\ \hline
$r_3=(1,0 ,0)$ & attractor & $s_2+s_3<s_1$ \\ \hline
$r_4=(0.3333,0.3333 ,0)$ & saddle & $s_3=0,s_1=s_2$ \\ \hline
$r_5=(0,1,0)$ & attractor & $s_3=0 ,s_1<s_2$ \\ \hline
$r_6\approx(0.3365 ,0.1682 ,0.1682)$ & saddle & The surface $s_1=s_2+s_3$ \\ \hline
The line $L=(0,s_2,1-s_2 )$ & attractor & $s_2+s_3>s_1$ \\ \hline
\end{tabular} \caption{ Fixed points, stability regimes, basins of attraction in the $\Gamma_4$ case. } \label {Table Gamma4}
\end{table} 

\section{CONCLUSIONS}\label{conclusions}

In this paper, we have tried to describe one of the most mysterious aspects of a human life, love, in  terms of complex system theories and nonlinear dynamics.
We have used the Verhulst-Lotka-Volterra prey-predator model  in order to  take into account competitive or cooperative attitudes between "agents". We have considered     3 agents ("lovers")  for establishing  the    minimal complexity   set of non linear differential equations,  and have  introduced different interaction matrices in order to mimick  some (well-known) situations in a 3 partner love story game.
We have  tried to answer some questions, like: What is the end of a love story in different scenarios? Under which condition will occur different situations? Which ones of these final situations are stable and under which conditions?

In this love triangle,
 we have further reduced the number of scenarios,  only considering that there is no   direct interaction (love or hate) between the lovers $A_2$ and $A_3$, yet keeping  $A_1$ in relation with  both neighboring partners. Therefore,   most of the time one of the rivals wins the love game and the other one will be completely ignored. However under some conditions,  as outlined in the main text, the game  does not   necessarily have any "winner";   there is a possible stable trio of lovers. Interestingly, this situation is very sensitive and is not stable. This means that it is hard to love several  persons simultaneously, - in this kind of model.

We have also considered the opposite of the previous  love triangle case, that is  $A_1$ is not interested in $A_2$ and $A_3$. As an example for this case consider the political relationship among three countries which are hostile together. We argued that in this case, in spite of the unfriendly direct interaction between the agents, there is always some agent looking for a friendly relationship; under special conditions, described in the previous sections, everybody demands friendly relationship with the others. But this situation is again not stable.
We have discussed two different scenarios in this so called "one-way love".  In the first scenario, the central lover $A_1$ is  interested by the two agents $A_2$ and $A_3$ but these do not love (care about)  $A_1$. The second scenario is the opposite:  the two lovers $A_2$ and $A_3$  love $A_1$, but the agent $A_1$ is not interested in them.    The analogy with  "public stars"  or "celebrity" has been mentioned in the main text.

In the first scenario, $A_1$ will eventually be deeply in love with the two others, but these two will not be fond of  $A_1$. Under special conditions in which $A_1$ starts the love game from a low amount of interest, the two agents $A_2$ and $A_3$ get somehow interested in the central lover $A_1$;   it seems they are appealing for more love and attention from the lover $A_1$ at the beginning of their love journey.
 
In the second scenario of one-way love, $A_1$ will not  be in love with the two other lovers, but these two lovers reach a balance point due to their competition in loving $A_1$. Just like in the first scenario, if the lovers $A_2$ and $A_3$ start the love game from a low amount of interest, the central agent will be interested in them and be looking for more love and attention,  right from the beginning of the love journey.

In our analysis, we have supposed that the strength of the interaction is  a binary variable, whatever  the  couples, and is time modulated in a self-organized, symmetric,  and preferential way through some Gaussian strength weight;  we have  also supposed that  the interaction term between $A_2$ and $A_3$  can be neglected. In order to achieve a deeper understanding it can be fruitful to reconsider these  constraints  in  future research. 
 It seems that it would be useful also to think about the evolution of love under "external fields",  and  about the dynamics of the system  over   time, allowing for  memory effects \cite{Jafari1aging,PRESIR,entropyjafariglass}.

\end{document}